\documentclass[final,5p,times,twocolumn]{elsarticle}

\usepackage{amssymb}
\usepackage{amsmath}
\usepackage{lipsum}
\usepackage{listings}
\usepackage{float}
\usepackage[colorlinks=true, linkcolor=blue, citecolor=blue, urlcolor=blue]{hyperref}
\usepackage{ulem}
\usepackage{xcolor}

\begin{document}

\begin{frontmatter}

\title{Identification and Online Monitoring of Experimental Measurement States via Cuscore Statistic}

\author[bh,rcnp]{Jichao Zhang}

\author[bh]{Baohua Sun\corref{cor1}}
\ead{bhsun@buaa.edu.cn}

\author[bh,rcnp]{Isao Tanihata}

\author[us]{Zilun Shen}

\cortext[cor1]{Corresponding author}

\affiliation[bh]{organization={School of Physics, Beihang University},
            city={Beijing},
            postcode={100191}, 
            country={China}}

\affiliation[rcnp]{organization={Research Center for Nuclear Physics (RCNP), Osaka University},
            city={Ibaraki Osaka},
            postcode={567-0047}, 
            country={Japan}}
            
\affiliation[us]{organization={Baruch College, City University of New York},
            city={New York},
            postcode={10010}, 
            country={USA}}

\begin{abstract}
We present a statistical method for detecting and analyzing state changes in experimental measurements using the Cuscore statistic and its special case, the Centred Cuscore statistic. 
These statistics are designed to identify deviations in detector responses using sequential hypothesis testing relative to a defined reference state.
Applications to charge-changing reaction experiments at the FRagment Separator facility at the GSI Helmholtz Centre for Heavy Ion Research, Germany, and the Second Radioactive Ion Beam Line in Lanzhou at the Institute of Modern Physics, China, demonstrate the ability of these tools to quantify state changes, identify the change point, and classify data segments based on measured states.
For long-term online monitoring, we use the exponentially weighted moving average to continuously update computations, enabling the detection of successive changes. 
This method supports both real-time and post-experiment diagnostics and provides a robust approach for enhancing data integrity and experimental control in nuclear physics and related fields.
\end{abstract}

\begin{keyword}

Cuscore statistic \sep Sequential probability ratio test \sep Nuclear experiment diagnostics \sep Online monitoring

\end{keyword}
\end{frontmatter}

\section{Introduction}

Advancements in nuclear physics increasingly depend on high-precision measurements, making the accuracy and reliability of experimental data a fundamental requirement for progress~\cite{he2023advances}. 
Detector systems, which provide the primary data for nuclear information extraction, are frequently affected by noise, instrumental drift, and subtle or difficult-to-detect malfunctions. 
As experimental setups grow in complexity\textemdash often involving multi-detector arrays\textemdash robust diagnostic tools are essential for evaluating detector performance, identifying and classifying experimental states, and enabling real-time monitoring.

In beamline-based nuclear physics experiments, the quality of the measured data is influenced by every component from the ion source to the final detection system.
In accelerator or storage-ring sections, as well as along the transport beamline, the magnetic fields define the beam optics and determine the particle transmission matrix~\cite{marie2008towards,droese2011investigation,li2023implementation}; even small magnetic drifts occurring during data acquisition can modify the transport conditions, resulting in an overlap of data from distinct experimental states within the same dataset, thereby introducing uncontrolled variations that can affect the final observables, such as in isochronous mass measurements~\cite{sun2010direct,tu2011precision,zhang2018isochronous}.
Meanwhile, the operational stability of detectors and their associated electronics governs the baseline, noise characteristics, and resolution of the recorded signals.
While advanced computational techniques, such as Bayesian algorithms and artificial intelligence optimizations, have been developed to model beam transport and improve accelerator stability~\cite{roussel2024bayesian,lopez2025ai}, a complementary approach is still required\textemdash one capable of diagnosing and confirming such effects directly from the measurement data itself.

A natural and fundamental viewpoint is that any change in the experimental state is inevitably encoded in the measured observables and is thus recoverable directly from the data stream.
This philosophy is fully aligned with long-standing practices in industrial monitoring, where system behaviour is assessed directly from the data.
In these fields, early practical implementations employed Shewhart charts, which directly compare each instantaneous observation with predefined control limits to detect abrupt departures from the baseline state~\cite{shewhart1931economic}.
Later, the cumulative sum (CUSUM) chart, which accumulates the successive departures of observations from a reference level to identify minor or gradually developing drifts with high sensitivity, was developed as statistical techniques advanced.\cite{page1954continuous,lucas1982combined}.

The cumulative score (Cuscore) statistic and its variant, the Centred Cuscore~\cite{box1992cumulative}, were originally developed for engineering process control and have since been widely used in financial trading~\cite{pole2011statistical}, offering a general statistical framework for characterizing deviations in sequential measurements.
Based on the statistical framework of the Sequential Probability Ratio Test (SPRT), Cuscore uses cumulative log-likelihood differences between the observed data and a reference state to quantify deviations.
Importantly, this method expands the ability to identify deviations against non-constant or structured reference functions, while retaining the classic CUSUM test as a special case (when the reference is a constant mean).
These features grant the method a high degree of flexibility, making it applicable to complex measurement environments.
In addition to its theoretical advantage, Cuscore also possesses several practical properties that make it particularly suitable for experimental diagnostics.
Because it is lightweight, visually interpretable, and does not require specialized contextual knowledge, the method can serve as an independent, rapid diagnostic tool. 
At the same time, its transparent statistical formulation makes it well suited to complement machine-learning approaches~\cite{boehnlein2022colloquium}\textemdash which typically require extensive data preparation and expert domain knowledge\textemdash by providing physically interpretable state information.
However, prior studies have mainly remained within the domain of theoretical statistics or industrial control. 
To date, the adaptation of the Cuscore framework for diagnosing detector performance or tracking state changes in experimental measurements, especially in the context of beamline-based nuclear physics, remains unexplored.

In this work, we introduce the Cuscore and Centred Cuscore methods to nuclear physics experiments for the first time, and develop a complete computational workflow that relies solely on measured detector observables to identify and quantify state changes.
By leveraging their ability to continuously track and accumulate variations in detector output over time or across events, we implement and demonstrate the framework in both retrospective offline analysis and real-time monitoring scenarios.
This approach offers a systematic means to detect both instabilities and significant deviations in measurement that may affect the integrity of the final results, and it provides diagnostic clues to help trace their possible origins.
These features make the method widely applicable across various experimental settings and research communities, with significant potential for incorporation into contemporary experimental nuclear physics workflows.

In this article, we first present the statistical model, detailing the procedures for setting calculation parameters and defining control boundaries, using simulated data as a baseline. 
Next, we demonstrate the practicality of the method through case studies involving charge-changing reaction measurements at the FRagment Separator (FRS) facility in the GSI Helmholtz Centre for Heavy Ion Research (GSI), Germany, and the Second Radioactive Ion Beam Line in Lanzhou (RIBLL2) in the Institute of Modern Physics (IMP), China. 
These examples illustrate the versatility and potential of Cuscore as a general-purpose tool for enhancing the reliability of complex experimental systems.

\section{Methods}

This study employs the Cuscore statistical framework to detect and characterize state changes in experimental measurements, that is, transitions where the detector output deviates from an expected reference behavior.

We begin by modeling the measured quantity $y_t$ at time (or event series) $t$ as
\begin{equation}\label{eq: yt define}
    y_t= T + a_t + f(t)  ~,
\end{equation}
where $T$ denotes the true value of the physical quantity, and $a_t$ represents the white noise of the detection system (assumed to be independent and identically distributed Gaussian noise, see~\ref{appendix: Cuscore} for details). $f(t)$ accounts for the state-dependent term expressed as $f(t) = \theta x(t)$, where $\theta$ is a constant indicating the amplitude of the state, and $x(t)$ describes the temporal behavior of the state as a function of time (or event series).

\subsection{Cuscore method}

The Cuscore method fundamentally arises from analyzing the likelihood function $L_t(\theta)$ of the residual sequence $a_t = y_t - T - f(t)$, to sequentially assess whether the measurement remains statistically consistent with a predefined reference state $\theta$. 
Throughout this paper, the reference state is generally defined as the baseline state $\theta_0$, representing the stable condition of the system, or as a state close to $\theta_0$ that characterizes the expected operating condition.

Taking the derivative of the log-likelihood $l_t(\theta)=\log L_t(\theta)$ with respect to $\theta$ at $\theta_0$ quantifies how deviations affect the likelihood of the observed residuals. The Cuscore $Q_t(\theta_0)$ is explicitly constructed on this principle:
\begin{equation}\label{eq: Q to l}
   Q_{t}(\theta_0) \propto     \left.\frac{\partial l_t}{\partial \theta}\right|_{\theta=\theta_{0}}=-\frac{1}{\sigma_a^{2}} \sum^{t}_{i=1} a_{i}(\theta_0) \left.\frac{\partial a_{i}}{\partial \theta}\right|_{\theta=\theta_{0}} ~,
\end{equation}
by defining
\begin{equation}\label{eq: D define}
      d_{t}(\theta_0)\equiv-\left.\frac{\partial a_{t}}{\partial \theta}\right|_{\theta=\theta_{0}} ~,
\end{equation}
and
\begin{equation}\label{eq: Q define 1}
      Q_{t}(\theta_0)\equiv \sum^{t}_{i=1} a_{i}(\theta_0) d_{i}(\theta_0)~\;,
\end{equation}
where the index $i$ runs over the sequence of measurements or events up to $t$. 
A detailed derivation is provided in \ref{appendix: Cuscore method}.

By accumulating statistical evidence over time, $Q_t(\theta)$ indicates whether the observed sequence diverges from a given reference state $\theta$. 
A sustained increase or decrease in the Cuscore indicates increasing inconsistency with the reference state, suggesting a possible state change. 
The specific expression of $Q_t$ depends on the chosen $f(t)$, which characterizes the type of signal.
Two representative types encountered in experimental measurements are introduced below.

\subsection{Constant signal}

We first introduce the case of a constant signal, in which the experimental state is expected to remain fixed over time. This corresponds to setting $x(t) = 1$, yielding $f(t) = \theta_0$ under the baseline condition.
When a state change occurs at the time $t_1$, the output term $f(t)$ deviates from its baseline value. This can be expressed as:
\begin{equation}\label{eq: drift at define}
   f(t)=\left\{\begin{matrix} 
  \theta_0~~~~~~~~~~~~t< t_1 \\  
  \theta_0+\theta'~~~~t \ge t_1
\end{matrix}\right. ~,
\end{equation}
where $\theta'$ denotes the shift of the signal.

Then we use the baseline state $\theta_0$ as the reference state for the Cuscore calculation to detect this change.
According to Eq.~\ref{eq: D define}, the corresponding detector function $d_t$ is:
\begin{equation}\label{eq: drft dt define}
    d_t=-\left.\frac{\partial a_{t}}{\partial \theta}\right|_{\theta=\theta_{0}}=1   ~.
\end{equation}

By applying Eq.~\ref{eq: Q define 1} under this formulation, the Cuscore $Q_{t}(\theta_0)$ becomes:
\begin{equation}\label{eq: Q define method}
\begin{aligned}
       Q_{t}(\theta_0) =& \sum^{t}_{i=1} a_{i}(\theta_0)d_{i}(\theta_0) \\
      =& \sum^{t}_{i=1} \left[y_i-T-\theta_0 \right] ~.
\end{aligned}
\end{equation}

\subsection{Periodic signal}

We also introduce the periodic signal, which satisfies $f(t) = f(t + C)$  for all $t$, where $C$ is the period. 
In this context, the true value $T$ remains constant, but the measured quantity exhibits a regular periodic deviation due to systematic or environmental effects. 
Such behavior can be captured by an appropriate function $f(t)$, which enables the Cuscore framework to account for and analyze periodic variations in experimental data.
A typical harmonic example is $f(t) = \theta \sin(\omega t)$, where $\omega$ is the angular frequency.
When the state of the harmonic signal amplitude changes at the time $t_1$, the corresponding $f(t)$ under the baseline condition is defined as:
\begin{equation}\label{eq: ramp at define}
   f(t)=\left\{\begin{matrix} 
  \theta_0 &\sin(\omega t)~~~~~~~~~~~~~~~~t< t_1 \\  
  (\theta_0+\theta')&\sin(\omega t)~~~~~~~~~~~~~~~~t \ge t_1
\end{matrix}\right. ~,
\end{equation}
where $\theta '$ is the same as defined in the constant signal. 

When setting baseline state $\theta_0$ as the Cuscore reference state, the corresponding detector function $d_t$ is:
\begin{equation}\label{eq: ramp dt define}
    d_t=-\left.\frac{\partial a_{t}}{\partial \theta}\right|_{\theta=\theta_{0}}=\sin(\omega t)   ~.
\end{equation}
The Cuscore $Q_{t}$ corresponding to this state change is given by:
\begin{equation}\label{eq: ramp Q define}
\begin{aligned}
       Q_{t}(\theta_0)&= \sum^{t}_{i=1} a_{i}(\theta_0)d_{i}(\theta_0) \\
       &= \sum^{t}_{i=1} \left[y_i-T-\theta_0 \sin(\omega i) \right]\sin(\omega i) ~.
\end{aligned}
\end{equation}

\subsection{Demonstration for Cuscore}
\label{Sec: Demonstration for Cuscore}

\begin{figure}[htbp] 
    \centering
    \includegraphics[width=1\linewidth]{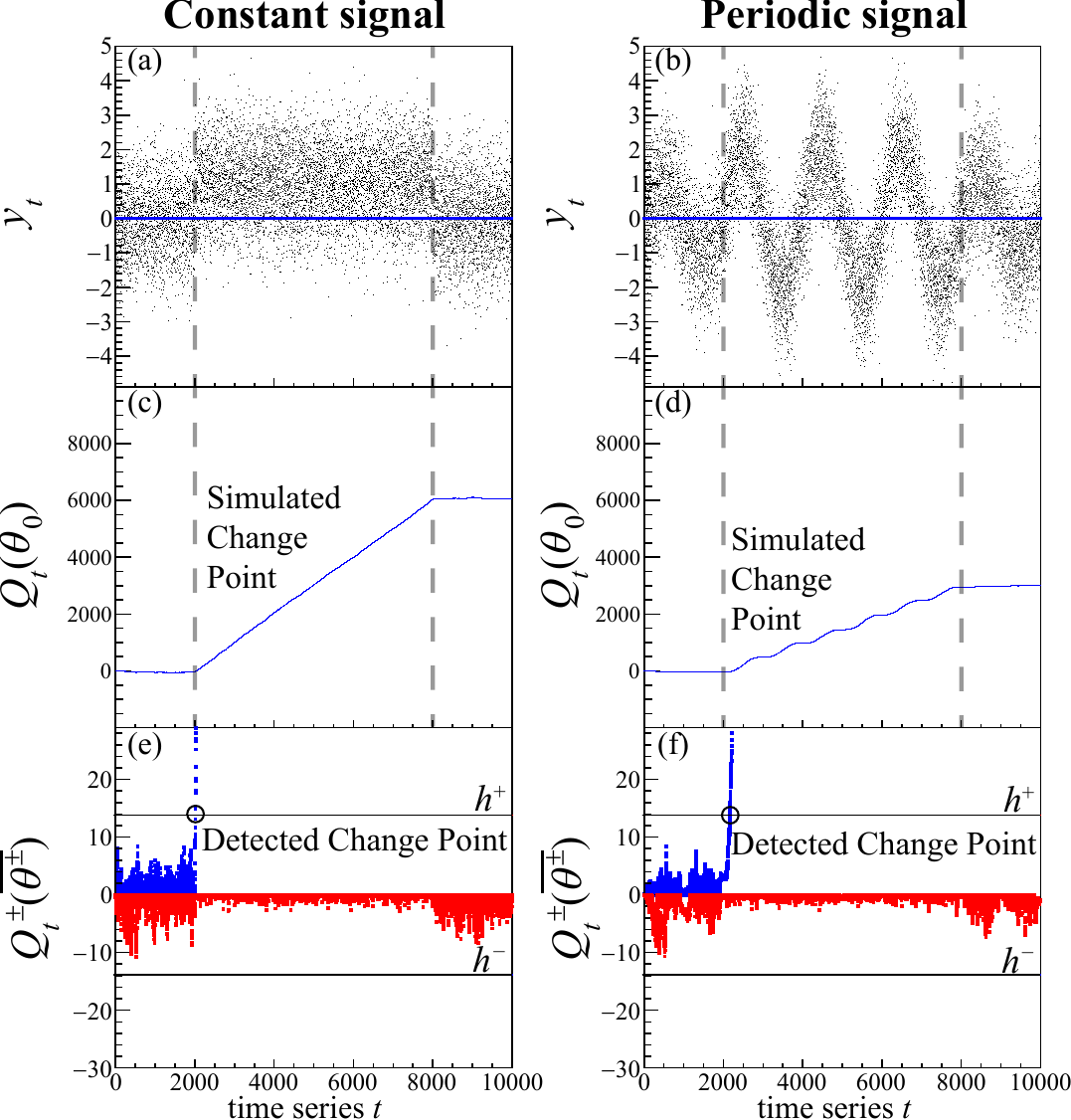}
    \caption{
    Schematic representation of the Cuscore method (see text for details).
    (a) and (b): Distribution of the simulated constant and periodic signals, $y_t$, over time series. 
    The blue solid line represents the true value, $T$, which is used to calculate the Cuscore. 
    (c) and (d): Cuscore of state $\theta_0$ for the two signals. The dashed lines indicate the simulated change point at $t$ = 2000 and 8000.
    (e) and (f): Two-directional Centred Cuscore for the two observables, showing the positive component $Q_t^+$ (upper branch, above zero; blue) and the negative component $Q_t^-$ (lower branch, below zero; red), together with the corresponding control boundaries $h^\pm$. Black circles along the threshold lines mark the detected change points.
}
    
    \label{fig: simulate}
\end{figure}

We illustrate the application of the Cuscore method by presenting two simulated signals: a constant signal and a periodic signal, as shown in Figures~\ref{fig: simulate}(a) and (b), respectively. 
Both are initially generated under baseline conditions defined by $\theta_0$, with the true value $T$ equal to 0 and the white noise term $a_t$ drawn from a Gaussian distribution of standard deviation $\sigma_a=1$. 
Specifically, the constant signal takes $f(t)=\theta_0=0$, while the periodic signal is given by $f(t)=\theta_0\sin(\omega t)$ with $\theta_0=1$ and $\omega=2\pi/2000$, i.e., $f(t)=\sin(2\pi/2000 \cdot t)$. 
In the simulation, each signal undergoes two state changes: at $t=2000$, the amplitude $\theta$ shifts from $\theta_0$ to 1 for the constant case and to 2 for the periodic case; at $t=8000$, both signals return to their original baseline states.
The magnitude of these changes matches the noise standard deviation, providing a representative test of the method’s sensitivity.

The corresponding Cuscore results are shown in Figures~\ref{fig: simulate}(c) and (d). 
In these calculations, the reference state is set to the baseline condition defined in the simulation, \textit{i.e.}, $\theta_0 = 0$ for constant and $\theta_0 = 1$ for periodic. 
Before the first imposed change at $t=2000$, the Cuscore $Q_t(\theta_0)$ in both signals remains nearly constant without noticeable increase or decrease trends, indicating that the system operates at the preset baseline state.
Following the state change at $t=2000$, $Q_t(\theta_0)$ in both cases rises rapidly, demonstrating that the method detects a positive deviation of the experimental state from the reference state $\theta_0$. 
The approximately constant rates of increase shown in Figure~\ref{fig: simulate}(c) indicate that the newly reached states are stable.
A comparable behavior is also evident in Figure~\ref{fig: simulate}(d), where the Cuscore maintains a similar overall slope despite appearing in a staircase pattern, which reflects the periodic nature of the input and does not contradict the underlying stability.
After $t=8000$, $Q_t(\theta_0)$ stops rising and stabilizes at a new constant value in both cases, indicating that another state change point has been detected and that the system has returned to its original baseline state. 
This demonstrates that Cuscore can accurately and rapidly identify state changes and their direction.

However, $Q_t(\theta_0)$ alone provides only a cumulative measure of divergence from the baseline state. 
At the onset of a potential shift, such accumulation does not permit a rigorous statistical distinction between a genuine state change and a rare fluctuation induced by noise. 
The absence of a formal decision threshold, therefore, limits its ability to provide a prompt and quantitative judgment, motivating the introduction of the Centred Cuscore.

\subsection{Centred Cuscore method}

To address the rigorous decision threshold, we introduce an additional reference state $\theta_1$ alongside the baseline state $\theta_0$. 
The difference $\theta_1 - \theta_0$  is selected to represent the maximum deviation that is still considered acceptable for the experimental state. 

We then employ the SPRT to compare the likelihoods under the two hypotheses $\theta_0$ and $\theta_1$~\cite{wald1992sequential}. 
When the log-likelihood ratio exceeds a preset threshold, it indicates that the data provide stronger support for $\theta_1$, implying that the state has deviated from $\theta_0$ to the boundary of the acceptable range. 
Formally, this decision rule is expressed as
\begin{equation}
    \log \frac{L_t(\theta_1)}{L_t(\theta_0)} = l_t(\theta_1) - l_t(\theta_0) > \ln\!\left(\frac{1}{\alpha}\right),
\end{equation}
where $\alpha$ is the significance level for the SPRT. 
When the inequality holds, we conclude\textemdash at significance level $\alpha$\textemdash that the experimental state has changed from $\theta_0$ to $\theta_1$.
In practical measurements, $\alpha$ can be estimated from the proportion of samples that trigger false alarms~\cite{dobben1968cumulative}.

To compute the log-likelihood ratio,
we expand it around the midpoint $\bar{\theta} = (\theta_0 + \theta_1)/2$.
In the general Cuscore framework, this yields the first-order Taylor approximation
\begin{equation}
    l_t(\theta_1) - l_t(\theta_0) \approx \frac{\theta_1-\theta_0}{ \sigma_{a}^{2}}\sum^{t}_{i=1}\left[a_{i}(\bar{\theta})d_{i}(\bar{\theta})\right] =\frac{\theta_1-\theta_0}{ \sigma_{a}^{2}}Q_{t}(\bar{\theta}) ~,
\end{equation}
with the full derivation given in Eqs.~\ref{eq: l1-l0}–\ref{eq: EQ}.
Here, we highlight that a key property of the specific linear model adopted in this work (Eqs.~\ref{eq: yt define}~and~\ref{eq: at define}) is that the residual $a_t(\theta)$ is a linear function of $\theta$. 
Hence its second derivative vanishes, $\partial^2 a_t(\theta) / \partial \theta^2 = 0$,  
which eliminates the second-order and all higher-order terms in the Taylor expansion.  
Consequently, for our linear model, the expression becomes an exact identity:
\begin{equation}
    l_t(\theta_1) - l_t(\theta_0)=\frac{\theta_1-\theta_0}{ \sigma_{a}^{2}}Q_{t}(\bar{\theta}) ~\;.
\end{equation}
This corresponds to evaluating the Cuscore at the centred state $\bar{\theta}$,
which naturally motivates calling it the Centred Cuscore method.
Therefore, when using the Centred Cuscore to evaluate state changes, the corresponding control boundary $h$ is given by:
\begin{equation} \label{EQ: h define}
Q_t(\bar{\theta}) >  \frac{\sigma_a^2 \ln(1/\alpha)}{\theta_1 - \theta_0}=h \;.
\end{equation}

In the model described above, the direction of the state change under consideration is predetermined, with $\theta_1 > \theta_0$ representing the target deviation.
However, in real experiments, the direction of state change is generally unknown.
To address this, it is necessary to introduce two reference states symmetrically placed on either side of the baseline $\theta_0$, allowing detection of deviations in both positive and negative directions.
Furthermore, to enhance sensitivity, each Cuscore branch continuously accumulates deviations in the direction it monitors, resetting to zero whenever the observed trend reverses.
This leads to the formulation of the two-directional Centred Cuscore~\cite{pham2023springer}, defined as follows:
\begin{equation}\label{eq: +-Q 1}
\begin{array}{c}
 Q_{t}^+(\overline{\theta^+})=\max\left[ 0, Q_{t-1}^+(\overline{\theta^+})+a_{t}(\overline{\theta^+})d_{t}(\overline{\theta^+}) \right]~,\\
 Q_{t}^-(\overline{\theta^-})=\min\left[ 0, Q_{t-1}^-(\overline{\theta^-})+a_{t}(\overline{\theta^-})d_{t}(\overline{\theta^-}) \right]~,\\
 Q_{0}^+(\overline{\theta^+})= Q_{0}^-(\overline{\theta^-})=0 ~.
\end{array}
\end{equation}
Here, upper $Q_{t}^+(\overline{\theta^+})$ corresponds to $\theta_1^+>\theta_0$, $\overline{\theta^+}=(\theta_1^++\theta_0)/2$. And lower $Q_{t}^-(\overline{\theta^-})$ corresponds to $\theta_1^-<\theta_0$, $\overline{\theta^-}=(\theta_1^-+\theta_0)/2$. 

When using the two-directional Centred Cuscore, since the values $Q_t^\pm(\overline{\theta^{\pm}})$ satisfy $|Q_t^\pm(\overline{\theta^{\pm}})| \geq | Q_t(\bar{\theta})|$, applying the same control threshold $h$ remains valid and enhances sensitivity to directional state changes, \textit{i.e.}:
\begin{equation} \label{EQ: h define 1}
|Q_{t}^{\pm}(\overline{\theta^{\pm}}) | > |h^{\pm}| = \frac{\sigma_a^2 \ln(1/\alpha)}{|\theta_1^{\pm} - \theta_0|} \;.
\end{equation}

\subsection{Demonstration for Centred Cuscore}

To demonstrate the practical use of the two-directional Centred Cuscore, we present the calculation results in Figures~\ref{fig: simulate}(e) and (f).
In both cases, we define the baseline states for Centred Cuscore calculations as $\theta_0 = 0$ and set the maximum acceptable deviation as $|\theta_1 - \theta_0| = 0.5$, equal to half the simulated signal state change.
This configuration enables effective monitoring of both positive and negative deviations.
The control boundary is computed from Eq.~\ref{EQ: h define 1} with a significance level $\alpha = 0.001$ for the SPRT, corresponding to a highly stringent criterion that allows us to identify a state change with strong statistical confidence when the threshold is exceeded, using the same noise level $\sigma_a = 1$ as employed in the signal generation.
A detailed discussion of the practical selection of these parameters, as well as their sensitivity in realistic applications, is provided in~\ref{appendix: Parameter in offline analysis}.

The results show the evolution of the Centred Cuscore over time, with the positive component $Q_t^+$ and the negative component $Q_t^-$.
Before $t=2000$, both $Q_{t}^+$ and $Q_{t}^-$ fluctuate within the control bounds $h^\pm$, indicating that the experimental state remains consistent with the baseline.
After the imposed state shift at $t = 2000$, $Q_t^+$ exceeds the control limit $+h$, thereby detecting the positive deviation.
A black circle along the threshold line indicates the time series of detected change points.
In Figure~\ref{fig: simulate}(f), for the periodic signal, the detection sensitivity is reduced compared to the constant signal, but the Centred Cuscore still crosses the control boundary after 200 time steps.
At $t=8000$, the system returns to its original state. Since this restored state does not deviate toward the negative reference direction, $Q_t^-$ does not exceed the control threshold.
These examples validate the sensitivity and directional capability of the Centred Cuscore method.
In our test, the simulated shift matched the white noise standard deviation, but the method remained sensitive to shifts as small as 1\% of the noise standard deviation.
This highlights its robustness in identifying weak deviations buried in noise.

In typical experimental scenarios, the observed quantity $y_t$, as defined in Eq.~\ref{eq: yt define}, remains statistically stable under the baseline state, where $T$ represents the true value of the measured quantity and $a_t$ denotes the detector resolution. 
In such cases, assuming a constant-state function $f(t)=\theta_0$ is generally sufficient to monitor experimental stability. 
Therefore, all subsequent data analyses in this work adopt a constant-state function in the Cuscore and Centred Cuscore calculations.

Nevertheless, when the functional form of the experimental state is known in advance\textemdash such as the harmonic periodic signal discussed earlier or other more complex time-dependent behaviors\textemdash it is possible to construct an appropriate state function $f(t)$ and incorporate it directly into the Cuscore framework. 
This enables the method to accommodate a wider range of experimental dynamics beyond constant states.

Importantly, Cuscore-based methods are not limited to detecting changes in amplitude. 
By designing the form of the observation $y_t$ accordingly, it is also possible to detect variations in signal distribution or other dynamic features, e.g., monitoring variance instead of amplitude as in Eq.~\ref{eq: yt define}. 
This flexibility makes Cuscore a powerful and adaptable tool for both offline data evaluation and real-time monitoring in precision experiments.

\section{Offline Data Analysis}
\label{Sec: Offline Data Analysis}

\begin{figure}[htbp] 
    \centering
    \includegraphics[width=0.7\linewidth]{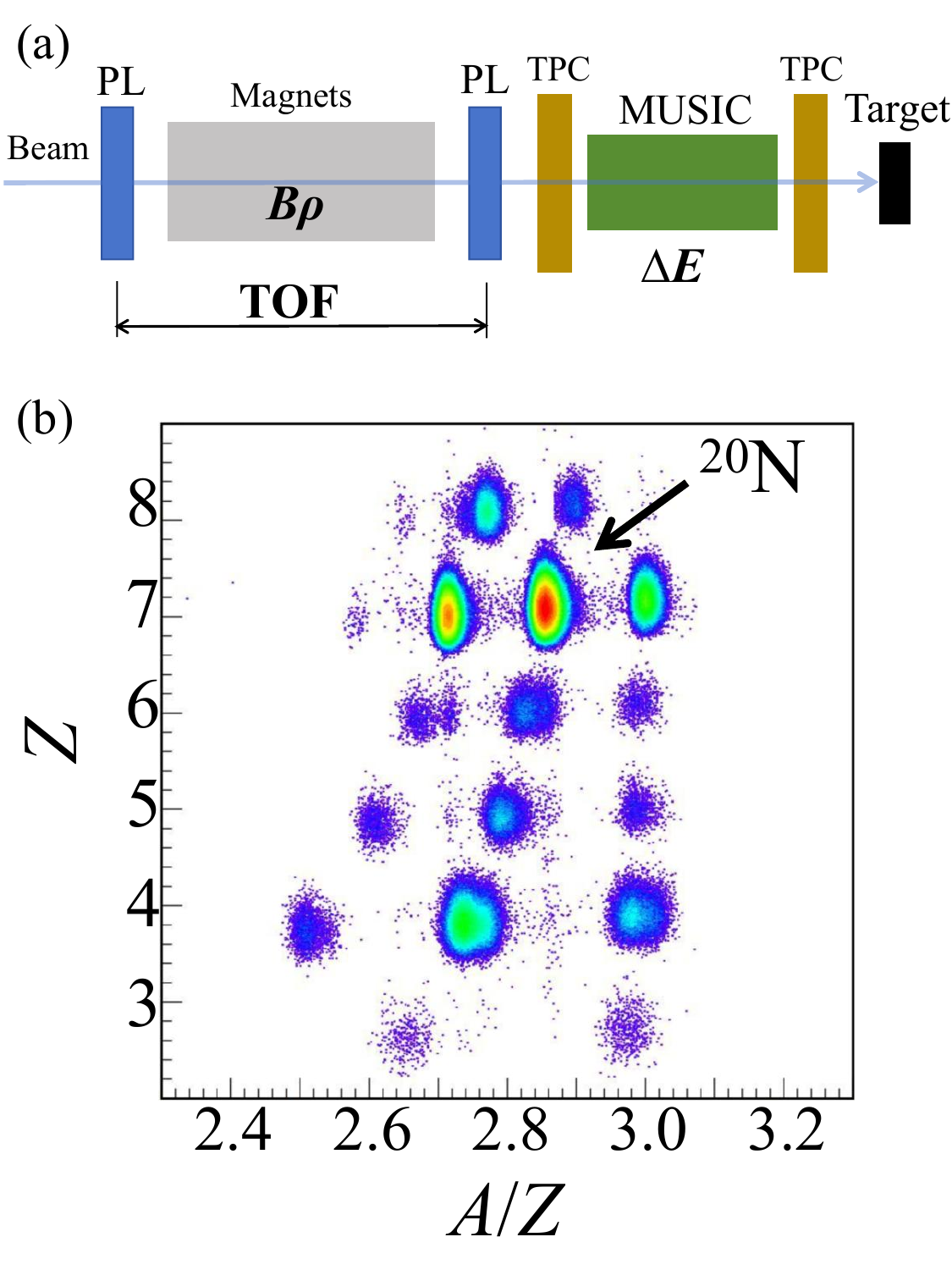}
    \caption{
(a) Schematic of the upstream experimental setup at the FRS, GSI for identifying incident particles before the reaction target.
(b) Corresponding particle identification spectrum of incident ions, with $^{20}$N highlighted by an arrow.}
    \label{fig: PID}
\end{figure}

Figure~\ref{fig: PID} presents an example from the charge-changing reaction measurements performed at the FRS at GSI, Germany. 
In this experiment, the magnetic rigidity ($B\rho$)-time of flight (TOF)-energy loss (${\Delta}E$) method is employed to determine the charge number ($Z$) and mass-to-charge ratio ($A/Z$) of incident particles, enabling particle identification (PID) before the target, as described in Refs.~\cite{zhang2024new,zhang2025cex}. 
The time-of-flight of ions along the beamline and the energy loss measured with the multisampling ion chamber (MUSIC) detector provide the key observables for particle identification. 
Both quantities are retrospectively analyzed using the Cuscore and Centred Cuscore statistical tools, identifying time points or intervals where state changes may have occurred during the measurement process.

\begin{figure}[htbp] 
\centering
    \includegraphics[width=1\linewidth]{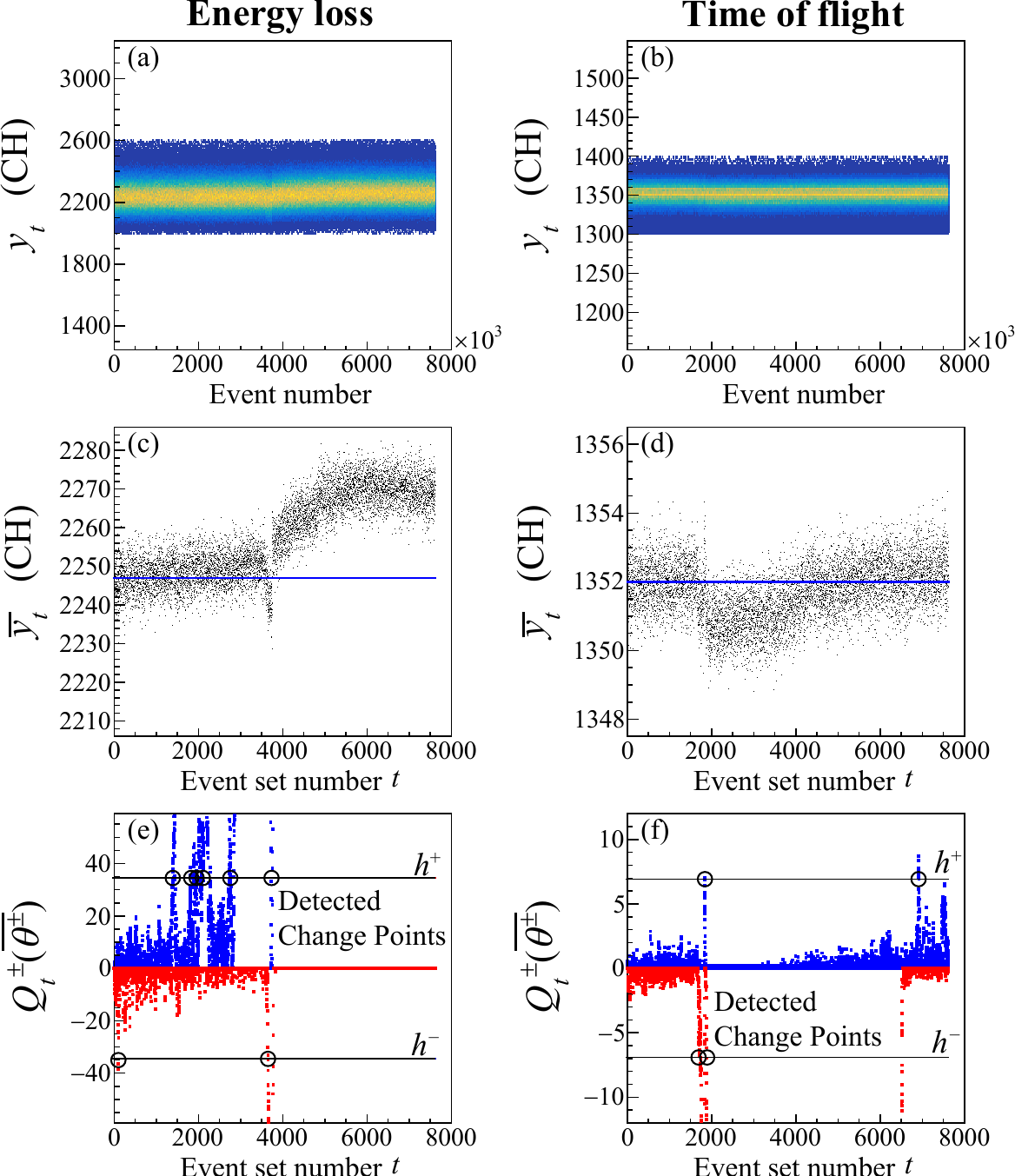}
    \caption{
     (a) and (b): Energy loss and time of flight for $^{20}$N before the reaction target.
     (c) and (d): Distribution of the mean values of energy loss and TOF of each subsequent 1000 events. The blue solid line represents the estimated true value $T$. 
     (e) and (f): Two-directional Centred Cuscore for energy loss and TOF signals, showing the positive component $Q_t^+$ (upper branch, above zero; blue) and the negative component $Q_t^-$ (lower branch, below zero; red), together with the corresponding control boundaries $h^\pm$. Black circles along the threshold lines mark the detected change points.
     }
    \label{fig: GSI}
\end{figure}

Figures~\ref{fig: GSI}(a) and (b) show the raw distributions of $\Delta E$ and TOF for incident $^{20}$N particles as a function of the event number, respectively.
Due to inherent fluctuations and structures, it is challenging to visually identify changes in the experimental state solely from the raw data. 
To improve interpretability, we calculate the average outputs of $\Delta E$ and TOF over every 1000 events, as shown in Figures~\ref{fig: GSI}(c) and (d). 
Such block-averaged values act as a scalar indicator of the current experimental state. 
By the central limit theorem~\cite{billingsley1986probability}, the block averages are approximately Gaussian.
Their central value can be used as the baseline true value $T$ in the Cuscore framework, and the fluctuations about this centre can be treated as an independent and identically distributed Gaussian noise term $a_t$ in Eq.~\ref{eq: yt define}.
The bin size 1000 is chosen as a representative order of magnitude, roughly corresponding to the typical number of events collected per second in the experiment.
A detailed discussion of how to select an appropriate bin size is provided in~\ref{appendix: Parameter in offline analysis}.
Such aggregation enables preliminary visual identification of experiment-state drifts.
In Figure~\ref{fig: GSI}(c), the average $\Delta E$ exhibits a transient drop around $t=3600$, followed by a gradual increase, and stabilizes only after $t=5000$.
In contrast, Figure~\ref{fig: GSI}(d) shows that the TOF starts decreasing as early as $t=2000$ and does not recover until approximately $t=5000$.
However, to accurately determine the timing and extent of these deviations, further quantitative analysis using the Centred Cuscore is necessary.

To implement the Centred Cuscore method, we first define the baseline state ($\theta_0 = 0$) by estimating the true detector outputs $T$.
Specifically, we use the average values from the initial 1000-event set in Figures~\ref{fig: GSI}(c) and (d) as estimates of $T$ for the true values of $\Delta E$ and TOF, respectively.
These values, indicated by the blue solid line in both figures, are used in the subsequent Cuscore calculations.
Furthermore, the maximum acceptable deviation from the baseline is chosen to be the intrinsic detector resolution, \textit{i.e.}, $|\theta_1^\pm - \theta_0| = \sigma_{\text{det}}$.
Figure~\ref{fig: GSI}(e) presents the two-direction Centred Cuscore $Q_{t}^\pm(\overline{\theta^\pm})$ for $\Delta E$, based on reference states $(\theta_1^\pm + \theta_0)/2$.
To determine the corresponding control boundaries $h^\pm$, the significance level $\alpha$ in Eq.~\ref{EQ: h define} is set to 0.001, and the noise standard deviation $\sigma_a$ is estimated as the standard deviation of the averaged $\Delta E$ values from the initial event sets (order numbers below 1000).
Similarly, Figure~\ref{fig: GSI}(f) shows the corresponding Centred Cuscore results for TOF, using the same analysis procedure as for $\Delta E$, and parameter selection based on Figure~\ref{fig: GSI}(d).

In Figure~\ref{fig: GSI}(e), the positive Cuscore crosses the control threshold at around the 1300th event and significantly fluctuates between the 2000th and 3600th event set.
These sustained signal fluctuations are difficult to discern visually in Figure~\ref{fig: GSI}(c), underscoring the enhanced sensitivity of the Cuscore approach.
Following event 3600, the negative Cuscore briefly exceeds the threshold, immediately followed by a sharp rise in the positive Cuscore that persists to the end of the event window, consistent with the visual trend observed in the $\Delta E$ signal.
In Figure~\ref{fig: GSI}(f), the negative Cuscore indicates a state change near the 2000th event, in agreement with the visual deviation in TOF seen in Figure~\ref{fig: GSI}(d).
Additionally, the positive Cuscore registers two brief excursions beyond the threshold around the 2000th and 7000th events, capturing transient state variations that are not as easily distinguishable by eye.

Given the demonstrated capability of the Centred Cuscore to detect state changes in experimental measurements, we recommend employing this method as an initial step in nuclear physics data analysis. 
By retrospectively identifying state transitions and their potential causes, one can categorize the data into distinct segments and treat each segment separately to improve the reliability of the results.
In the present example, once the incident particle ($^{20}$N) is selected, both the measured TOF and $\Delta E$ reflect its velocity upon reaching the detectors. 
Therefore, simultaneous changes in these two observables likely indicate variations in the particle’s transport conditions\textemdash such as subtle drifts in the magnetic field settings\textemdash as seen in the simultaneous positive change around the 2000th event set and the correlated variation between the 3000th and 5000th event sets. 
In contrast, isolated triggers in either $\Delta E$ or TOF are more plausibly attributed to transient instabilities in the corresponding detector or electronics systems, such as the brief drop in $\Delta E$ around the 100th and 3600th event sets or the TOF negative shift near the 2000th.
From a statistical standpoint, even with a stringent significance level ($\alpha = 0.001$, i.e., a nominal false-alarm probability of order $10^{-3}$), isolated threshold crossings cannot be rigorously excluded as chance events. 
From a data-analysis perspective, however, crossings that persist over several consecutive bins are exceedingly unlikely to arise from random fluctuations alone and are therefore interpreted as evidence of a real change in the experimental state.
After $t = 5000$, both observables stabilize in a new state, suggesting that the system has entered a new stable condition. 
Based on this analysis, we extract cross sections separately from data before $t = 2000$ and after $t = 5000$, while excluding events near the 100th, 1300th, and 7000th that triggered Cuscore alarms. 
The two resulting segments yield consistent cross sections; however, the result obtained from the unsegmented dataset differs by approximately 2~mb for $^{20}$N, as reported in Ref.~\cite{zhang2024new}.
This underscores the effectiveness of the approach used. 

As demonstrated in our example, the Cuscore method not only detects state changes in individual observables but also enables further inference by examining the correlation between different quantities. 
For instance, simultaneous variations in both energy loss and time-of-flight measurements suggest a possible drift in the beamline magnetic fields. 
Such cross-checks of logical consistency among observables broaden the scope of system checking and enhance the overall reliability of the experimental analysis. 
It is essential to note that several configurable parameters are involved in implementing the model, including the event grouping size, the significance level, and the selection of the reference state. 
These parameters may be selected according to the guidelines in~\ref{appendix: Guidelines for Parameter Selections} or tuned to the specific characteristics and requirements of the experimental setup.

\section{Online Monitoring}
\label{Sec: Online Monitoring}

The Cuscore and Centred Cuscore methods can be applied in practice to evaluate the experimental state at each moment $t$, enabling their use in online monitoring systems. 
When the true value $T$ in Eq.~\ref{eq: yt define} is estimated based on the initial experimental baseline state $\theta_0$, the time of the first state change can be detected promptly. 
However, monitoring subsequent state changes becomes difficult unless the system returns to the initial reference state.
For example, as shown in Figures~\ref{fig: simulate}(e) and (f), the Centred Cuscore effectively detects the initial positive change given by $Q_{t}^+$ at $t=2000$, but it fails to identify the second negetive change by $Q_{t}^-$ at $t=8000$, because $Q_{t}^-$ is constantly detecting negative deviations from the initial reference state.
In many experimental scenarios, it is not necessary to maintain consistency with the initial state; rather, it is sufficient for the experimental system to remain stable in any acceptable state. 
To enable long-term monitoring across multiple state changes, the baseline state must adapt by dynamically updating the true value $T$, so that the Cuscore method remains sensitive to deviations from the evolving experimental condition, rather than a fixed state.

Various models exist for real-time estimation of the true value, including moving-average models and Kalman filters~\cite{kalman1960new}. 
In this work, we employ the exponentially weighted moving average (EWMA) method~\cite{brown2004smoothing}, which provides computational simplicity and adjustable sensitivity. 
This flexibility makes it well-suited to match different monitoring requirements.
The EWMA is defined recursively as follows: 
\begin{equation}\label{eq: EWMA}
   y_t^{\text{EWMA}}=\left\{\begin{matrix} 
   y_t~~~~~~~~~~~~~~~~~~~~~~~~~~~~~~~~~~~~~~~~~t=0  \\  
  \lambda y_{t-1}^{\text{EWMA}}+(1-\lambda)y_{t-1}~~~~~~~~~~t > 0
\end{matrix}\right. ~.
\end{equation}
where $y_t$ is the observation at time $t$, $y_t^{\text{EWMA}}$ is the smoothed EWMA estimate, and $\lambda \in [0,1]$ is the discount factor. 
The discount factor determines how quickly older observations become less relevant to the current estimate. 
When $\lambda$ is near 1, the estimate highly depends on past contributions. When it is near 0, it is closer to the previous observation $y_{t-1}$.

\begin{figure}
    \centering
    \includegraphics[width=1\linewidth]{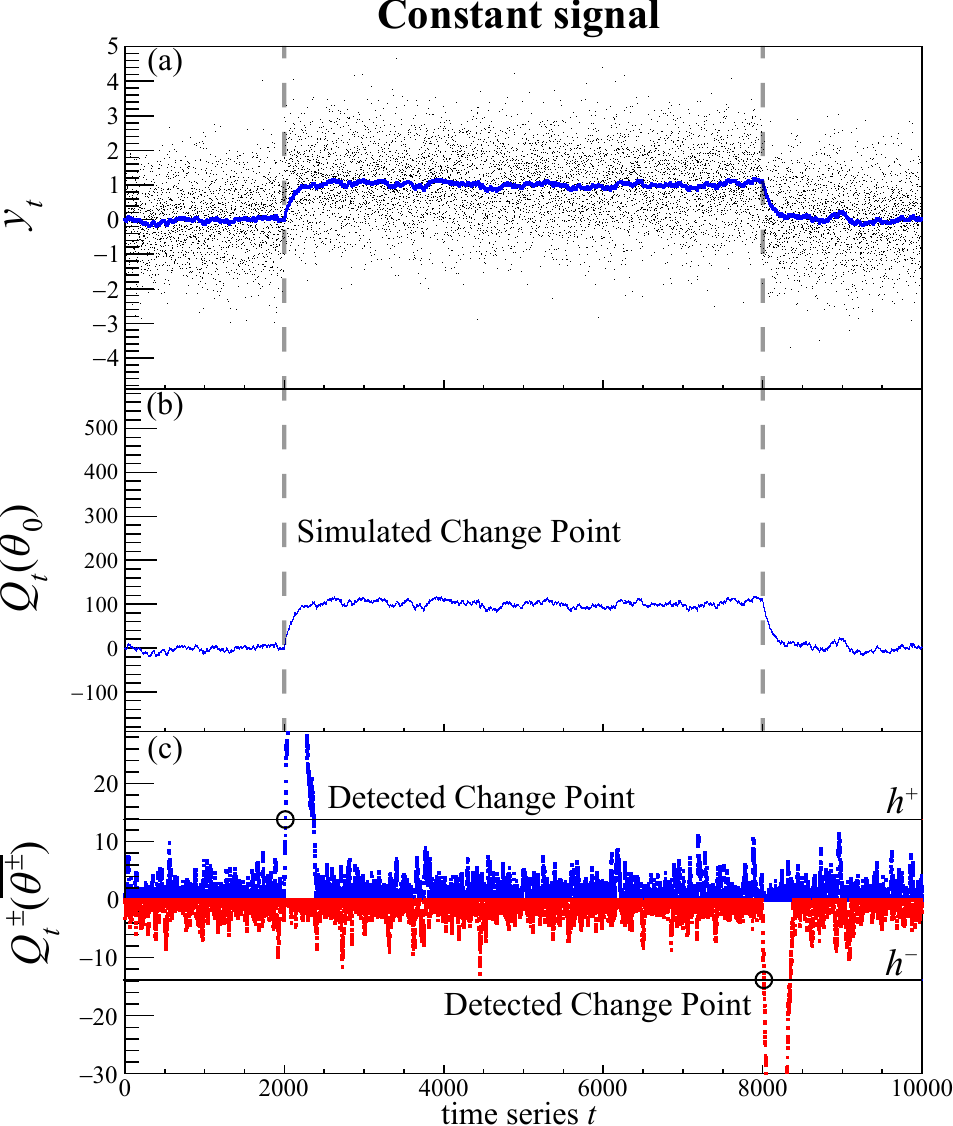}
    \caption{
    (a): The distribution of the simulated constant signal, $y_t$, over time series. The real-time true-value series $T_t$ is generated using the EWMA method (blue solid line). 
    (b): The Cuscore of state $\theta_0$. 
    (c): Two-directional Centred Cuscore, showing the positive component $Q_t^+$ (upper branch, above zero; blue) and the negative component $Q_t^-$ (lower branch, below zero; red), together with the corresponding control boundaries $h^\pm$. Black circles along the threshold lines mark the detected change points.
    }
    \label{fig: EWMA}
\end{figure}

Figure~\ref{fig: EWMA} demonstrates the application of the EWMA-enhanced Centred Cuscore on the same constant-type dataset used in Figure~\ref{fig: simulate}. 
Here, the true value $T$ in Eq.~\ref{eq: yt define} is replaced by the real-time true value series $T_t = y_t^{\text{EWMA}}$, as defined in Eq.~\ref{eq: EWMA} with $\lambda = 0.99$.
The derivation of $\lambda$ is detailed in~\ref{appendix: Parameter in online monitoring}. 
This allows the Cuscore to track state changes adaptively while preserving sensitivity to deviations.
As shown in Figure~\ref{fig: EWMA}(a), the EWMA estimate $y_t^{\text{EWMA}}$ closely follows the evolving experimental state. 
Figure~\ref{fig: EWMA}(b) shows the corresponding Cuscore, where a flat trajectory indicates a stable experimental state.
In contrast, when a fixed $T$ is used, as in Figure~\ref{fig: simulate}(c), the stability is indicated by a linear trend in $Q_t(\theta_0)$. 
The Centred Cuscore serves as a real-time alarm system and successfully identifies all the change points. 
As shown in Figure~\ref{fig: EWMA}(c), both the positive drift at $t = 2000$ and the negative drift at $t=8000$ are clearly and promptly detected.

\begin{figure}
    \centering
    \includegraphics[width=1\linewidth]{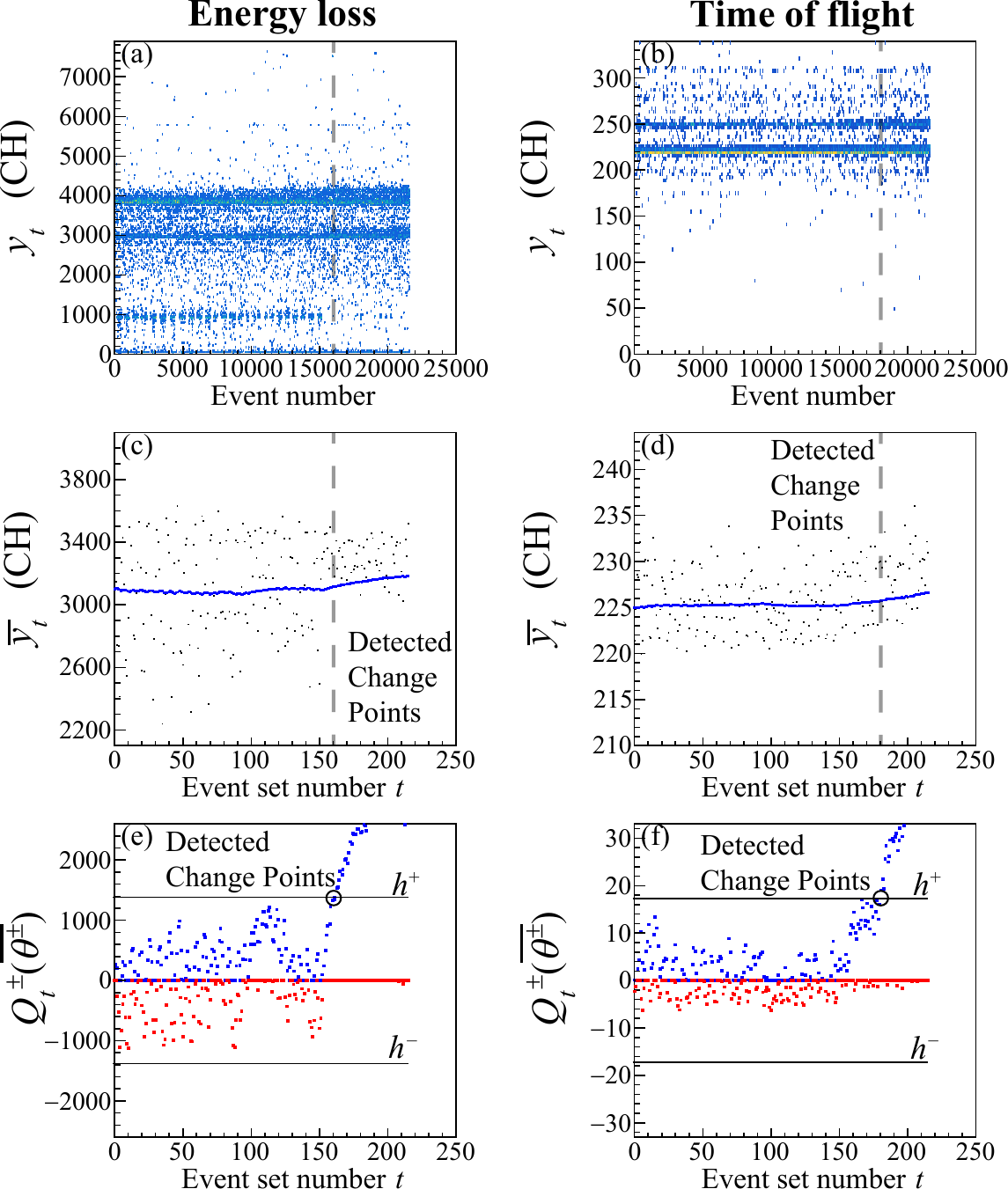}
    \caption{
    Same as Figure.~\ref{fig: GSI} but for the online data at IMP. See the text for details. 
    (a) and (b): Energy loss and time of flight before the reaction target. 
    (c) and (d): Distribution of the mean values of energy loss and TOF for 100 events concerning the order of event sets. The blue solid line represents the estimated true value $T_t$ given by the EWMA method. 
    (e) and (f): Two-directional Centred Cuscore for energy loss and TOF signals, showing the positive component $Q_t^+$ (upper branch, above zero; blue) and the negative component $Q_t^-$ (lower branch, below zero; red), together with the corresponding control boundaries $h^\pm$. Black circles along the threshold lines mark the detected change points.
    }
    \label{fig: IMP}
\end{figure}

This approach has been successfully implemented for online monitoring of the charge-changing reaction measurement at the RIBLL2 beamline at IMP, Lanzhou~\cite{wang2023charge,xu2025full}. 
Figure~\ref{fig: IMP} presents such an application, which is nearly identical to the GSI experiment in the previous section, employing the $B\rho$-TOF-$\Delta E$ method to identify secondary particles based on the energy loss and time-of-flight measurements.

Figures~\ref{fig: IMP}(a) and (b) show the raw energy loss and TOF data accumulated for a fixed $B\rho$ setting. 
The events include all incident particle types and represent the overall data distribution. 
A visible change in the energy loss distribution is observed after the 15,000th event in Figure~\ref{fig: IMP}(a), while no such apparent shift is seen in the TOF data in Figure~\ref{fig: IMP}(b).
To monitor potential global state changes during the experiment, the online Cuscore method is applied to the continuous data stream.
Figures~\ref{fig: IMP}(c) and (d) show the distribution of the mean values for each grouped event set, along with EWMA's estimate of the true value with $\lambda=0.99$ (blue solid line). 
Due to the low counting rate of secondary ions in the experiment, each subsequent 100 events has been grouped into one event set.
After this aggregation, a clear upward shift in energy loss emerges beyond the 150th event set, and the EWMA trend rises accordingly—indicating a possible change in experimental conditions.
Meanwhile, a gradual upward trend in TOF is also observed, though it remains difficult to determine whether this constitutes a true state change from visual inspection alone.
Figures~\ref{fig: IMP}(e) and (f) show the Centred Cuscore monitoring plot provided by the online data acquisition system.
All parameters are chosen as the guidelines in \ref{appendix: Guidelines for Parameter Selections}.
The experimental state is observed to change after the event set at 160th and fails to recover thereafter. 
Specifically, in Figure~\ref{fig: IMP}(e), the Centred Cuscore for energy loss rapidly exceeds the control boundary following event set 160th. 
Simultaneously, the TOF Cuscore in Figure~\ref{fig: IMP}(f) approaches the threshold at the same point and eventually surpasses it by event set 180th.
As in the GSI case, the concurrent change in both observables suggests a common underlying cause—most likely a shift in the magnetic field affecting beam transport.
Subsequent investigation confirmed this hypothesis: a fluorescent target had inadvertently dropped into the beamline, disturbing the beam trajectory.

The above examples demonstrate that the Cuscore method enables both offline analysis and real-time monitoring of experimental state changes. 
The EWMA-enhanced variant further extends its applicability to long-term trend tracking. 
By analyzing multiple observables in parallel, the method can also uncover latent causes behind detected changes\textemdash for instance, correlated shifts in energy loss and TOF suggest underlying magnetic field drift.
Building on these capabilities, the Cuscore framework can be integrated into experimental control systems as a reliable alert mechanism, thereby improving the stability, adaptability, and robustness of complex experimental setups.\\

\section{Discussion on the Limitations of Cuscore}

In its current form, the Cuscore framework is based on the modeling choice for the observable $y_t$ and a set of underlying statistical assumptions. 
These design choices are well-suited to identify the signal-shift state change considered in this work, but they also introduce several practical limitations for real experimental applications. 
For clarity, the main constraints and their implications are summarised below.

In this work, we model the observable $y_t$ as in Eq.~\ref{eq: yt define} and decompose it into an expected baseline value $T$, a white-noise term $a_t$, and a state term $\theta x(t)$ that is linear in the experimental state $\theta$, which provides a rigorous statistical basis.
First, $T$, $a_t$, and related parameters for the reference state must be obtained either from a baseline segment of the measured data or from prior estimates (see~\ref{appendix: Guidelines for Parameter Selections}).
And to ensure that $a_t$ is well approximated by independent and identically distributed Gaussian noise, the measurements are binned, and the bin averages are analysed, so that the central limit theorem applies; this introduces a minimum bin size (of order 50 events in the present context) and thus a minimum effective resolution in event (or time) index. 
In addition, the present $y_t$ model is tailored to detect deviations in signal amplitude (mean shifts). 
If the experimental state is expected to preserve the mean while changing the distribution shape, the model must be extended.
Within the Cuscore framework, this can be achieved by applying Eq.~\ref{eq: yt define} to a higher-order statistic of the data, such as the variance.

In the EWMA-enhanced online monitoring, a relatively large discount factor $\lambda$ is used to stabilise the baseline estimate, thereby slowing the response to abrupt changes, and brief state changes may appear weakened in the online Cuscore trace.
For this reason, we recommend confirming such a state change by a retrospective offline analysis within the Cuscore framework.

An inherent limitation of Cuscore (and, more generally, of statistical tests) is that it flags when and in which direction the experimental state changes, but it cannot determine the physical cause on its own.
Interpreting a detected change, therefore, requires experiment-specific knowledge of the accelerator, beamline, and detector systems.

\section{Summary}

In this work, we introduce the Cuscore and Centred Cuscore statistical tools for identifying and classifying state changes in experimental measurements, with a particular application to nuclear physics experiments. 
These methods support both offline data analysis and online monitoring by capturing subtle shifts in the measured quantities.

The Cuscore method is derived from the derivative of the log-likelihood with respect to the state parameter, and the centred Cuscore incorporates the sequential probability ratio test to establish decision thresholds for identifying significant deviations.
Applied to charge-changing reaction data from the GSI FRS and IMP RIBLL2, these methods successfully identified the onset of experimental instabilities and enabled classification of data into reliable segments for subsequent analysis.
Moreover, the evolution patterns of the monitored observables provide clues to the possible sources of these instabilities: in the GSI data, the concurrent variations in $\Delta E$ and TOF may reflect a drift in the beam-transport magnetic field.

For real-time online monitoring, we incorporated the EWMA model to dynamically estimate and update the true values of observed quantities.
This allows the reference state in the Cuscore framework to evolve with the experimental conditions, enabling the detection of deviations relative to each newly established stable state.
As a result, the method supports robust long-term monitoring across successive state changes throughout the experiment.
In the IMP experiment, the real-time application of the method revealed an abnormal response, later confirmed to have originated from a foreign object entering the beamline.

The proposed framework is adaptable, computationally efficient, and requires only minor assumptions about the structure of experimental data. 
By focusing on key observables and linking their variations to underlying experimental conditions, the Cuscore approach not only detects state changes but also helps interpret their physical origins.
These features make it a powerful diagnostic tool for improving experimental reliability, guiding real-time operational decisions, and ensuring data quality in complex measurement environments.

Looking ahead, we will further develop the potential of the Cuscore framework by coupling it more tightly to the existing data-acquisition systems and deploying it across a broader range of measurement platforms.
On the theoretical side, we plan to extend the present formulation beyond the linear state model considered here, so that more complex, nonlinear state evolutions—such as oscillatory behaviour (e.g. frequency drifts) and changes in the underlying distribution—can be monitored within a unified statistical framework.
Ultimately, these advances will facilitate the integration of Cuscore-based diagnostics into automated monitoring and control systems for future nuclear and accelerator experiments.

\section*{Acknowledgements}
The authors gratefully acknowledge the accelerator and FRS technical staff at the GSI for their dedicated technical support in preparing and maintaining the experimental setup.
We also thank the members of the experimental collaboration involved in the GSI campaign for their valuable contributions during data acquisition.
Similarly, we express our gratitude to the accelerator and RIBLL2 beamline staff at IMP, as well as to the participants of the corresponding experimental team, for their assistance and collaboration.
This work was partly supported by the National Natural Science Foundation of China (No. 12325506) and the 111 Center (Grant No. B20065).

\appendix 
\section{Cuscore statistic}\label{appendix: Cuscore}
\subsection{Cuscore}\label{appendix: Cuscore method}
We begin by formalizing the statistical model used to describe the detector output in this work. We defined the observed value $y_t$ at time (or event index) $t$ as~\cite{box1992cumulative}:
\begin{equation}\label{eq: yt again}
y_t = T + a_t + f(t) = T + a_t + \theta x_t ~,
\end{equation}
where $T$ denotes the true value of the measured quantity, and the sequence $a_t$ represents the white noise component of the detection system, modeled as an independent and identically distributed Gaussian variable with zero mean and variance $\sigma_a^2$ by the standard normal theory.
The function $f(t) = \theta x_t$ captures the experimental state as a time-dependent perturbation, where $x_t$ is a known function of time (or event order), and $\theta$ characterizes the state amplitude.

We assume that the experiment proceeds normally at the defined baseline state $\theta = \theta_0$. 
Under this assumption, the residuals $a_t$ can be described by:
\begin{equation}\label{eq: at define}
a_t(\theta_0) = y_t - T - \theta_0 x_t ~,
\end{equation}
and the likelihood of observing a sequence ${a_1, a_2, \dots, a_t}$ is given by:
\begin{equation}
    L_{t}(\theta_0)=\prod^{t}_{i=1} \exp \left\{-\frac{1}{2 \sigma_{a}^{2}}\left[a_{i}^{2}(\theta_0)\right]\right\}  ~,
\end{equation}
After taking the logarithm, the likelihood is transformed into:
\begin{equation}
    l_{t}(\theta_0)=\log{L_{t}(\theta_0)}=-\frac{1}{2\sigma_a^2}\sum^t_{i=1} a_{i}^2(\theta_0)  ~.
\end{equation}
The sensitivity of the likelihood to changes in $\theta$ is captured by the derivative of the log-likelihood:
\begin{equation}
    \left.\frac{\partial l_t}{\partial \theta}\right|_{\theta=\theta_{0}}=-\frac{1}{\sigma_a^{2}} \sum^{t}_{i=1} a_{i}(\theta_0) \left.\frac{\partial a_{i}}{\partial \theta}\right|_{\theta=\theta_{0}} ~.
\end{equation}
We set:
\begin{equation}\label{eq: dt define}
    -\left.\frac{\partial a_{t}}{\partial \theta}\right|_{\theta=\theta_{0}}=d_{t}(\theta_0) ~,
\end{equation}
then
\begin{equation}
    \left.\frac{\partial l_t}{\partial \theta}\right|_{\theta=\theta_{0}}=\frac{1}{\sigma_a^{2}} \sum^{t}_{i=1} a_{i}(\theta_0) d_{i}(\theta_0) ~.
\end{equation}
Thus we define
\begin{equation}\label{eq: Q define}
   Q_{t}(\theta_0)= \sum^{t}_{i=1} a_{i}(\theta_0) d_{i}(\theta_0)={\sigma_a^{2}}\left.\frac{\partial l_t}{\partial \theta}\right|_{\theta=\theta_{0}}~.
\end{equation}
Here $Q_{t}(\theta_0)$ is the Cuscore when the reference state $\theta=\theta_0$.

For an arbitrary state $\theta$ at the time of the experiment, expanding $a_{t}$ at $\theta_0$, there are
\begin{equation}\label{eq: at expand}
   a_{t}(\theta)=a_{t}(\theta_0) - (\theta-\theta_0)d_{t}(\theta_0)~.
\end{equation}
Note that this is an exact derivation of the linear model in Eq.~\ref{eq: at define}, not an approximation. 
This linearity is the foundational property that subsequently ensures the exactness of the Centred Cuscore derivation (see~\ref{appendix: Centred Cuscore}).

When the experimental state deviates from the ideal state $\theta_0$, that deviation is added to $a_t(\theta)$ by the increment of the vector $d_{t}(\theta_0)$, resulting in a noticeable change seen in Eq.~\ref{eq: Q define}.
Thus, the Cuscore statistic represented by Eq.~\ref{eq: Q define} actually looks for this particular change continuously over the sequence \textit{t}.

\section{Centred Cuscore statistic}
\subsection{Centred Cuscore}
\label{appendix: Centred Cuscore}

To quantitatively monitor changes in the experimental state, we introduce a second reference state $\theta = \theta_1$, in addition to the baseline state $\theta = \theta_0$. 
The difference $\theta_1 - \theta_0$ is selected to represent the maximum deviation that is still considered acceptable for the experimental state. 
This allows us to formulate a sequential probability ratio test for detecting if the state of the experiment is changing from $\theta_0$ to $\theta_1$, 
\begin{equation}\label{eq: l1-l0}
    l_{t}(\theta_1)-l_{t}(\theta_0)=-\frac{1}{2 \sigma_{a}^{2}}\sum^{t}_{i=1}\left[a_{i}^{2}(\theta_1)-a_{i}^{2}(\theta_0)\right]  ~.
\end{equation}

We consider a centred state $\bar{\theta}=(\theta_1+\theta_0)/2$. 
Expanding $a_{t}$ at $\bar{\theta}$, we get the following:
\begin{equation}\label{eq: at expand2}
   a_{t}(\theta)\approx a_{t}(\bar{\theta}) - (\theta-\bar{\theta})d_{t}(\bar{\theta}) ~.
\end{equation}
For a general nonlinear model, Eq.~\eqref{eq: at expand2} would represent a
first-order Taylor approximation, with higher-order terms omitted.  
However, in the specific context of this work, the underlying signal model defined in Eq.~\ref{eq: at define} is linear in the parameter $\theta$, implying
\begin{equation}
\frac{\partial^2 a_t(\theta)}{\partial\theta^2}=0.
\end{equation}
By the Lagrange form of the Taylor remainder, the second-order and all higher-order terms therefore vanish identically, and Eq.~\eqref{eq: at expand2} becomes an exact relation rather than an approximation.  
This exactness ensures that the subsequent derivation of the Centred Cuscore statistic is mathematically rigorous within the linear framework considered here.

And the Eq.~\ref{eq: l1-l0} becomes:
\begin{equation}\label{eq: l1-l0 2}
\begin{aligned}
    &l_{t}(\theta_1)-l_{t}(\theta_0) = \\
    &-\frac{1}{2 \sigma_{a}^{2}}\sum^{t}_{i=1} \left\{ \left[ a_{i}(\bar{\theta}) - (\theta_1-\bar{\theta})d_{i}(\bar{\theta}) \right]^2   -\left[ a_{i}(\bar{\theta}) - (\theta_0-\bar{\theta})d_{i}(\bar{\theta}) \right]^2 \right\}   \\
    &=\frac{\theta_1-\theta_0}{ \sigma_{a}^{2}}\sum^{t}_{i=1}\left[a_{i}(\bar{\theta})d_{i}(\bar{\theta})\right] =\frac{\theta_1-\theta_0}{ \sigma_{a}^{2}}Q_{t}(\bar{\theta}) ~.
\end{aligned}
\end{equation}
The right-hand side is exactly Cuscore \textit{Q} when the testing state is assumed as $\theta=\bar{\theta}$, which is also called the \textit{Centred Cuscore}.

Since the least squares estimator of $\theta-\bar{\theta}$ is $\Sigma a_{i}(\bar{\theta})d_{i}(\bar{\theta})/\Sigma d^2_{i}(\bar{\theta})$, hence the expectation of $Q_{t}(\bar{\theta})$ is
\begin{equation}\label{eq: EQ}
   E[Q_{t}(\bar{\theta})]=(\theta-\bar{\theta})\sum^{t}_{i=1} d^2_{i}(\bar{\theta}) ~.
\end{equation}
We see that the Centred Cuscore $Q_{t}(\bar{\theta})$ will increase when $\theta>\bar{\theta}$ and decrease when $\theta<\bar{\theta}$.

\subsection{Significance tests}

In a standard SPRT, the ``null hypothesis'' assumes that the system remains in its original, controlled state (i.e., no change in $\theta$ has occurred). 
A decision is made only when sufficient evidence accumulates to reject this hypothesis.
By contrast, the Cuscore test can be interpreted as a logically reversed SPRT~\cite{johnson1961simple}: it assumes the control state may already have changed and continuously evaluates whether there is sufficient evidence to indicate a change has occurred. 
At every step, it calculates the likelihood of a deviation from the control state, making it highly sensitive to detecting small, gradual changes over time.

From this perspective, we decide that the experiment transitions from state $\theta_0$ to state $\theta_1$ if
\begin{equation}\label{eq: l1-l0 3}
   l_{t}(\theta_1)-l_{t}(\theta_0)> \ln (1/\alpha) ~.
\end{equation}
where $\alpha$ is the significance level for the test. In Cuscore, it can also be seen as a rough approximation of the proportion of samples that trigger false alarms~\cite{dobben1968cumulative}.
By joining Eq.~\ref{eq: l1-l0 2} and Eq.~\ref{eq: l1-l0 3}, we get the Cuscore test allowable bounds are then:
\begin{equation}
   Q_{t}(\bar{\theta})>\frac{\sigma_{a}^2\ln (1/\alpha)}{\theta_1-\theta_0}=h~,
\end{equation}
where $\sigma_{a}$ is the variance of the white noise sequence, which can be regarded as the variance of the measurement.
This parameter should be determined from simulations or preliminary analyses of the experimental data.
A state change is considered to occur when the Centred Cuscore exceeds the boundary, i.e., $Q_{t}(\bar{\theta})>h$ or $Q_{t}(\bar{\theta})<-h$.

In real experiments, the direction of mutation of the experimental state is uncertain. Therefore, we use two different directions of the Centred Cuscore~\cite{pham2023springer}. That is, upper $Q_{t}^+(\overline{\theta^+})$ corresponds to $\theta_1^+>\theta_0$, $\overline{\theta^+}=(\theta_1^++\theta_0)/2$. And lower $ Q_{t}^-(\overline{\theta^+})$ corresponds to $\theta_1^-<\theta_0$, $\overline{\theta^-}=(\theta_1^-+\theta_0)/2$. 
Both start at zero and accumulate unilaterally, i.e.
\begin{equation}\label{eq: +-Q}
\begin{array}{c}
 Q_{t}^+(\overline{\theta^+})=\max\left[ 0, Q_{t-1}^+(\overline{\theta^+})+a_{t}(\overline{\theta^+})d_{t}(\overline{\theta^+}) \right]~,\\
 Q_{t}^-(\overline{\theta^-})=\min\left[ 0, Q_{t-1}^-(\overline{\theta^-})+a_{t}(\overline{\theta^-})d_{t}(\overline{\theta^-}) \right]~,\\
 Q_{0}^+(\overline{\theta^+})= Q_{0}^-(\overline{\theta^-})=0 ~.
\end{array}
\end{equation}
At this time, the control conditions are:
\begin{equation}
  Q_{t}^+(\overline{\theta^+}) >h^+=\frac{\sigma_{a}^2\ln (1/\alpha)}{\theta_1^+-\theta_0}~ \mathrm{,~~and~~} Q_{t}^-(\overline{\theta^-}) <h^-=\frac{\sigma_{a}^2\ln (1/\alpha)}{\theta_1^--\theta_0}~.
\end{equation}
When the experiment is stable over a long period, a one-sided Cuscore test is preferable because it filters out the partial effects of small drifts in $Q_t$, thus maintaining the effectiveness of the monitoring.

\section{Guidelines for Parameter Selection}\label{appendix: Guidelines for Parameter Selections}

For the offline and online applications discussed in Sec.~\ref{Sec: Offline Data Analysis} and~\ref{Sec: Online Monitoring}, the Cuscore and Centred-Cuscore statistics involve several user-defined parameters that control the trade-off between sensitivity and robustness. 
This section outlines their roles, practical selection criteria, and the impact of their variation.

\subsection{Parameter in offline analysis}\label{appendix: Parameter in offline analysis}

\subsubsection*{Bin size:}

For experimental data, we first bin the measurements into fixed intervals and use the bin averages as the state indicator.
By the central limit theorem~\cite{billingsley1986probability}, these averages are approximately Gaussian, so the fluctuations about the baseline can be treated as an independent and identically distributed Gaussian noise term $a_t$, consistent with the Cuscore assumptions.
Under approximately constant beam intensity, this is equivalent to averaging over a short time interval. 
A practical choice is to set the bin size to the number of events collected within about one second.
To balance temporal resolution and statistical robustness, the bin size should not exceed the beam extraction duration per spill, while an empirical lower bound in typical nuclear physics experiments is 50 events per bin to ensure approximate normality.

\subsubsection*{True value $T$:}
In the offline analysis, the model requires a true value $T$ representing the expected value in the baseline state $\theta_0$. 
When this quantity is known a priori—as in the simulations of Sec.~\ref{Sec: Demonstration for Cuscore}—it can be used directly. 
When $T$ is unknown, it should be estimated from data acquired under stable operating conditions. 
As an example, in the GSI case, we use the mean of the initial 1000-event sets, taken from a baseline state period.
If $T$ is set too high, upward shifts tend to be underestimated and downward shifts over-reported; if it is set too low, the opposite bias occurs.

\subsubsection*{Noise standard deviation $\sigma_a$:}
The quantity $\sigma_a$ is determined by the standard deviation of the bin-averaged value of the observation. 
It is typically estimated together with the reference true value $T$ using data segments judged stable.
Because $\sigma_a^2$ is proportional to the control boundary $h$, an inaccurate estimate directly affects the sensitivity of state change detection. 
Since the value of $\sigma_a$ depends on the chosen bin size, it must be re-evaluated whenever the binning configuration is modified.

\subsubsection*{Maximum acceptable deviation $|\theta_1 - \theta_0|$:}
This parameter specifies the maximum deviation from the baseline state $\theta_0$ that remains acceptable under normal operating conditions.
In nuclear physics experiments, a practical choice is to set $|\theta_1 - \theta_0|$ comparable to the intrinsic detector resolution (e.g., \ $|\theta_1 - \theta_0| = \sigma_{\mathrm{det}}$), ensuring that the method reacts to physically meaningful deviations.
If the analysed observable does not have a well-defined intrinsic resolution, we recommend using a rough scale of $\sigma_a/2$.
If $|\theta_1 - \theta_0|$ is chosen too small, the procedure inevitably produces many false alarms driven by random noise.
If it is chosen too large, only relatively strong drifts can cross the corresponding boundary, and weaker changes that are nevertheless meaningful may not be identified.

\subsubsection*{Significance level $\alpha$:}

The significance level $\alpha$ specifies the tolerated false-alarm probability, i.e.\ the probability that a threshold crossing arises from random fluctuations rather than a genuine state change.
For example, $\alpha = 0.01$ implies that approximately 1\% of alarms may be false. 
Given that nuclear physics instrumentation is generally stable and that false triggers may lead to unnecessary data segmentation or misinterpretation, we recommend using a stringent value of $\alpha = 0.001$, with an upper admissible limit of 0.01.

\subsection{Parameter in online monitoring}
\label{appendix: Parameter in online monitoring}

\subsubsection*{EWMA model:} 
The EWMA model is adopted for the estimated true value $T_t$ because it is computationally inexpensive and does not require storing historical measurements.
These two practical advantages are crucial in settings where large data volumes must be continuously monitored, and the bandwidth of the data-acquisition system may be limited.
Although the Kalman filter is a more general estimator~\cite{kalman1960new}, in the special case relevant here—where we aim to detect small deviations around the baseline state—it could be reduced to the EWMA form. 
Its additional complexity and requirement for precise noise modelling therefore offer no practical advantage.
Machine-learning predictors, despite their utility in high-dimensional pattern analysis, are not recommended for baseline estimation.
Their data-driven outputs require substantial training data and are typically less stable, which means they lack the interpretability necessary for Cuscore’s statistical test.

More broadly, machine-learning techniques can, in principle, be applied to anomaly-state detection, particularly when many detector channels must be analysed jointly, and complex multivariate correlations are involved~\cite{boehnlein2022colloquium}. 
Their strengths, however, are less aligned with the diagnostic tasks considered here, in which the relevant quantities are low-dimensional, well-defined, and changed around a stable baseline. 
Data-driven models may introduce additional variability through training-set dependence or overfitting, thereby reducing transferability and obscuring genuine state changes.
For such applications, the Cuscore+EWMA framework provides sufficient sensitivity and interpretability while avoiding the need for model training. 
These considerations make it a practical and reliable choice for the experimental state identification tasks examined in this study, whereas machine-learning methods remain more suitable for complex, high-dimensional diagnostic settings.

\subsubsection*{Discount factor $\lambda$:} 

The EWMA model is used to generate the baseline estimate $T_t$ for the online Cuscore monitor.  
This estimator must satisfy two requiremen: it should evolve slowly enough to provide a stable reference against which the Cuscore statistic can accumulate evidence of a genuine deviation; and its own fluctuations must remain sufficiently small so as not to interfere with the subsequent SPRT decision.  
These considerations imply that the discount factor $\lambda$ cannot be chosen too small.

To quantify this constraint, we begin with the steady-state variance of the EWMA estimator.  
When the input is white noise with variance $\sigma_a^2$, the variance of $y_t^{\mathrm{EWMA}}$ is approximately~\cite{crowder1989design}
\begin{equation}
\label{eq: Var}
\operatorname{Var}\!\left(y_t^{\mathrm{EWMA}}\right) 
\approx \frac{1-\lambda}{1+\lambda}\,\sigma_a^2.
\end{equation}
To ensure that EWMA-induced fluctuations do not cause false triggers in the Cuscore test, we require the $K$-standard-deviation bound to remain below the Cuscore control boundary:
\begin{equation}
\label{eq: K}
K\sqrt{\operatorname{Var}\!\left(y_t^{\mathrm{EWMA}}\right)} 
\;\le\; 
h = \frac{\sigma_a^2 \ln(1/\alpha)}{|\theta_1 - \theta_0|},
\end{equation}
where $K$ is a safety factor and $\alpha$ is the significance level of the Cuscore's SPRT.  
Substituting Eq.~\ref{eq: Var} and solving for $\lambda$ yields
\begin{equation}
\label{eq: lambda}
\lambda \ge 
\frac{
1 - \left( \dfrac{|\theta_1 - \theta_0|}{K\sigma_a\sqrt{\ln(1/\alpha)}} \right)^2
}{
1 + \left( \dfrac{|\theta_1 - \theta_0|}{K\sigma_a\sqrt{\ln(1/\alpha)}} \right)^2
}.
\end{equation}

For our monitoring system, we maintain consistency with the Cuscore test parameters:
The safety factor is set to $K = 3.29$, which is the two-tailed critical value obtained by looking up the standard normal distribution table for 
a confidence level of $99.9\%$ (or equivalently, a significance level of $0.001$), thereby matching the significance level $\alpha$ of the Cuscore's SPRT.
And following the discussion above, we set the maximum acceptable deviation to $|\theta_1 - \theta_0| = \sigma_a/2$.  
Substituting these values into Eq.~\ref{eq: lambda}:
\begin{equation}
\lambda \geq \frac{1 - \left( \frac{0.5}{3.29 \cdot \sqrt{\ln(1000)}} \right)^2}{1 + \left( \frac{0.5}{3.29 \cdot \sqrt{\ln(1000)}} \right)^2} \approx 0.99.
\end{equation}

Thus, the theoretically permissible minimum discount factor is $\lambda_{\min} \approx 0.99$, representing an appropriate balance between baseline stability and responsiveness.  
Choosing a larger $\lambda$ value would further suppress EWMA's own fluctuations but at the cost of an excessively slow response to genuine drifts, providing no additional benefit relative to our reliability criterion.  
If different significance levels $\alpha$ or different sensitivities $|\theta_1 - \theta_0|$ are desired, the corresponding $\lambda$ should be recalculated using Eq.~\ref{eq: lambda}.

\bibliographystyle{elsarticle-num}

\begin{thebibliography}{10}
\expandafter\ifx\csname url\endcsname\relax
  \def\url#1{\texttt{#1}}\fi
\expandafter\ifx\csname urlprefix\endcsname\relax\def\urlprefix{URL }\fi
\expandafter\ifx\csname href\endcsname\relax
  \def\href#1#2{#2} \def\path#1{#1}\fi

\bibitem{he2023advances}
R.~He, X.-Y. Niu, Y.~Wang, H.-W. Liang, H.-B. Liu, Y.~Tian, H.-L. Zhang, C.-J. Zou, Z.-Y. Liu, Y.-L. Zhang, et~al., Advances in nuclear detection and readout techniques, Nuclear Science and Techniques 34~(12) (2023) 205.

\bibitem{marie2008towards}
M.~Marie-Jeanne, J.~Alonso, K.~Blaum, S.~Djekic, M.~Dworschak, U.~Hager, A.~Herlert, S.~Nagy, R.~Savreux, L.~Schweikhard, et~al., Towards a magnetic field stabilization at isoltrap for high-accuracy mass measurements on exotic nuclides, Nuclear Instruments and Methods in Physics Research Section A: Accelerators, Spectrometers, Detectors and Associated Equipment 587~(2-3) (2008) 464--473.

\bibitem{droese2011investigation}
C.~Droese, M.~Block, M.~Dworschak, S.~Eliseev, E.~M. Ramirez, D.~Nesterenko, L.~Schweikhard, Investigation of the magnetic field fluctuation and implementation of a temperature and pressure stabilization at shiptrap, Nuclear Instruments and Methods in Physics Research Section A: Accelerators, Spectrometers, Detectors and Associated Equipment 632~(1) (2011) 157--163.

\bibitem{li2023implementation}
Y.~Li, X.~Wang, H.~Zhang, Q.~Wang, J.~Li, N.~Zhou, T.~Gao, X.~Yan, Y.~Huang, D.~Gao, Implementation of a dipole magnet power supply control system to improve magnetic field stability at the csre storage ring facility for precision mass measurement, Nuclear Instruments and Methods in Physics Research Section A: Accelerators, Spectrometers, Detectors and Associated Equipment 1049 (2023) 168108.

\bibitem{sun2010direct}
B.~Sun, R.~Kn{\"o}bel, H.~Geissel, Y.~A. Litvinov, P.~Walker, K.~Blaum, F.~Bosch, D.~Boutin, C.~Brandau, L.~Chen, et~al., Direct measurement of the 4.6 mev isomer in stored bare 133sb ions, Physics Letters B 688~(4-5) (2010) 294--297.

\bibitem{tu2011precision}
X.~Tu, M.~Wang, Y.~A. Litvinov, Y.~Zhang, H.~Xu, Z.~Sun, G.~Audi, K.~Blaum, C.~Du, W.~Huang, et~al., Precision isochronous mass measurements at the storage ring csre in lanzhou, Nuclear Instruments and Methods in Physics Research Section A: Accelerators, Spectrometers, Detectors and Associated Equipment 654~(1) (2011) 213--218.

\bibitem{zhang2018isochronous}
Y.~Zhang, P.~Zhang, X.~Zhou, M.~Wang, Y.~A. Litvinov, H.~Xu, X.~Xu, P.~Shuai, Y.~Lam, R.~Chen, et~al., Isochronous mass measurements of t z=- 1 fp-shell nuclei from projectile fragmentation of ni 58, Physical Review C 98~(1) (2018) 014319.

\bibitem{roussel2024bayesian}
R.~Roussel, A.~L. Edelen, T.~Boltz, D.~Kennedy, Z.~Zhang, F.~Ji, X.~Huang, D.~Ratner, A.~S. Garcia, C.~Xu, et~al., Bayesian optimization algorithms for accelerator physics, Physical review accelerators and beams 27~(8) (2024) 084801.

\bibitem{lopez2025ai}
S.~Lopez-Caceres, D.~Santiago-Gonzalez, Ai-assisted transport of radioactive ion beams, Physical Review Accelerators and Beams 28~(7) (2025) 072802.

\bibitem{shewhart1931economic}
W.~Shewhart, Economic control of quality of manufactured product (1931).

\bibitem{page1954continuous}
E.~S. Page, Continuous inspection schemes, Biometrika 41~(1/2) (1954) 100--115.

\bibitem{lucas1982combined}
J.~M. Lucas, Combined shewhart-cusum quality control schemes, Journal of quality technology 14~(2) (1982) 51--59.

\bibitem{box1992cumulative}
G.~Box, J.~Ram{\'\i}rez, Cumulative score charts, Quality and Reliability Engineering International 8~(1) (1992) 17--27.

\bibitem{pole2011statistical}
A.~Pole, Statistical arbitrage: algorithmic trading insights and techniques, John Wiley \& Sons, 2011.

\bibitem{boehnlein2022colloquium}
A.~Boehnlein, M.~Diefenthaler, N.~Sato, M.~Schram, V.~Ziegler, C.~Fanelli, M.~Hjorth-Jensen, T.~Horn, M.~P. Kuchera, D.~Lee, et~al., Colloquium: Machine learning in nuclear physics, Reviews of modern physics 94~(3) (2022) 031003.

\bibitem{wald1992sequential}
A.~Wald, Sequential tests of statistical hypotheses, in: Breakthroughs in statistics: Foundations and basic theory, Springer, 1992, pp. 256--298.

\bibitem{dobben1968cumulative}
C.~v. Dobben~de Bruyn, Cumulative sum tests: theory and practice, Gri n's Statistical Monographs and Courses 24 (1968).

\bibitem{pham2023springer}
H.~Pham, Springer handbook of engineering statistics, Springer Nature, 2023.

\bibitem{zhang2024new}
J.-C. Zhang, B.-H. Sun, I.~Tanihata, R.~Kanungo, C.~Scheidenberger, S.~Terashima, F.~Wang, F.~Ameil, J.~Atkinson, Y.~Ayyad, et~al., A new approach for deducing rms proton radii from charge-changing reactions of neutron-rich nuclei and the reaction-target dependence, Science Bulletin 69~(11) (2024) 1647--1652.

\bibitem{zhang2025cex}
J.-C. Zhang, B.-H. Sun, I.~Tanihata, S.~Terashima, F.~Wang, R.~Kanungo, C.~Scheidenberger, F.~Ameil, J.~Atkinson, Y.~Ayyad, et~al., Charge pickup reaction cross section for neutron-rich p-shell isotopes at 900 a mev, Physical Review X 15~(3) (2025) 031004.

\bibitem{billingsley1986probability}
P.~Billingsley, \href{https://books.google.co.jp/books?id=Q2IPAQAAMAAJ}{Probability and Measure}, Wiley Series in Probability and Statistics, Wiley, 1986.
\newline\urlprefix\url{https://books.google.co.jp/books?id=Q2IPAQAAMAAJ}

\bibitem{kalman1960new}
R.~E. Kalman, A new approach to linear filtering and prediction problems (1960).

\bibitem{brown2004smoothing}
R.~G. Brown, Smoothing, forecasting and prediction of discrete time series, Courier Corporation, 2004.

\bibitem{wang2023charge}
C.-J. Wang, G.~Guo, H.~J. Ong, Y.-N. Song, B.-H. Sun, I.~Tanihata, S.~Terashima, X.-L. Wei, J.-Y. Xu, X.-D. Xu, et~al., Charge-changing cross section measurements of 300 mev/nucleon 28si on carbon and data analysis, Chinese Physics C 47~(8) (2023) 084001.

\bibitem{xu2025full}
X.-D. Xu, Y.~Zheng, Z.-Y. Sun, Y.-N. Song, B.-H. Sun, S.~Terashima, C.-J. Wang, G.~Guo, G.-S. Li, X.-L. Wei, et~al., Full realization of the ribll2 separator at the hirfl-csr facility, Science bulletin (2025) S2095--9273.

\bibitem{johnson1961simple}
N.~L. Johnson, A simple theoretical approach to cumulative sum control charts, Journal of the American Statistical Association 56~(296) (1961) 835--840.

\bibitem{crowder1989design}
S.~V. Crowder, Design of exponentially weighted moving average schemes, Journal of Quality technology 21~(3) (1989) 155--162.

\end{thebibliography}

\end{document}